\documentclass[aps,floatfix,amsmath,twocolumn,superscriptaddress,notitlepage,nofootinbib,longbibliography]{revtex4-2}

\usepackage{graphicx}
\usepackage{dcolumn}
\usepackage{bm}
\usepackage{xcolor}
\usepackage{hyperref}
\usepackage[ruled]{algorithm2e}
\SetKwFor{For}{for (}{) $\lbrace$}{$\rbrace$}
\usepackage{amssymb}
\usepackage{graphicx}
\usepackage{placeins}
\usepackage{braket} 
\usepackage{amsthm}

\usepackage{dsfont}
\usepackage{wrapfig}

\usepackage[utf8]{inputenc}
\usepackage[T1]{fontenc}
\usepackage{float}
\usepackage{mathtools}
\usepackage{geometry}
\usepackage{xcolor}

\usepackage{bbm}

\usepackage{enumitem}
\usepackage[normalem]{ulem}

\hypersetup{
    colorlinks=true,
    linkcolor=blue,
    citecolor=blue,
    urlcolor=blue
}

\newcommand{\Tr}{\mathrm{Tr}}
\def\ketbra#1#2{\mathinner{|{#1}\rangle\!\langle{#2}|}}

\def\dket#1{\mathinner{|{#1}\rangle\!\rangle}}

\def\dketbroj#1{\mathinner{|{#1}\rangle\!\rangle\!\langle\!\langle{#1}|}}

\newcommand{\id}{\mathds{1}}

\usepackage{amsfonts}

\newtheoremstyle{mythmstyle}
  {10pt}{10pt}{\itshape}{}{\bfseries}{.}{.5em}{}
\theoremstyle{mythmstyle}

\theoremstyle{definition}

\makeatletter 
    
\renewcommand\onecolumngrid{
\do@columngrid{one}{\@ne}%
\def\set@footnotewidth{\onecolumngrid}
\def\footnoterule{\kern-6pt\hrule width 1.5in\kern6pt}%
}

\renewcommand\twocolumngrid{
        \def\footnoterule{
        \dimen@\skip\footins\divide\dimen@\thr@@
        \kern-\dimen@\hrule width.5in\kern\dimen@}
        \do@columngrid{mlt}{\tw@}
}%

\makeatother    
\definecolor{cool_green}{rgb}{0.0, 0.5, 0.0}

\begin{document}

\title{How many systems can be dephased \\ before the quantum switch becomes causally definite?}

\author{Yassine Benhaj}
\affiliation{Université Grenoble Alpes, CNRS, Grenoble INP, Institut Néel, 38000 Grenoble, France}
\author{Kuntal Sengupta}
\affiliation{Université Grenoble Alpes, CNRS, Grenoble INP, Institut Néel, 38000 Grenoble, France}
\affiliation{Université Grenoble Alpes, Inria, 38000 Grenoble, France}
\author{Cyril Branciard}
\affiliation{Université Grenoble Alpes, CNRS, Grenoble INP, Institut Néel, 38000 Grenoble, France}

\date{\today}

\begin{abstract}

Quantum processes with indefinite causal order -- so-called causally nonseparable processes -- can exhibit various advantages over quantum circuits with a fixed or a well-defined causal structure. A natural question is how much nonclassicality is required for a process to display causal nonseparability.
Here we address this by investigating how many systems can be dephased (or decohered) before this property vanishes.
First, for bipartite processes with open past and future we show that if all systems are dephased, or if only the future system is kept undephased, then the process becomes causally separable. However, if any single system other than the future system remains undephased, then there exist processes that retain causal nonseparability. Next, we demonstrate a similar behaviour in the multipartite case, when restricted to the physically motivated class of quantum circuits with quantum control (QC-QCs). Namely, dephasing all systems or keeping only the future system undephased renders any QC-QC causally separable; while causal nonseparability can persist if any non-future system is left undephased.

\end{abstract}

\maketitle

Higher order operations within quantum theory feature processes that allow operations to take place without a well-defined causal order~\cite{Chiribella2013,OCB2012}. This feature, often referred to as \textit{indefinite causal order} or \textit{causal nonseparability}, has recently received much attention, both in quantum foundations, in the understanding of causation, and in its ability to provide leverage in information processing tasks over scenarios where all operations occur in definite order. 
A natural enquiry is to investigate how much quantumness, or non-classicality is required for a process to feature indefinite causal order.
In this work, we study this problem from the standpoint of the number of systems that can be dephased until a process becomes causally separable.

\medskip

It is known from Ref.~\cite{OCB2012} that for bipartite processes without global past $(P)$ and future $(F)$, dephasing all systems makes any process causally separable. This result was generalized in~\cite{Baumann2016Appearance}, where it was shown that it suffices to dephase the input systems of the two parties to imply causal separability; while as soon as just one of the parties' input systems is coherent, one can construct causally nonseparable processes~\cite{OCB2012}.
However, the question for bipartite processes with global past $(P)$ and future $(F)$ systems has so far remained open.

A paradigmatic, well-studied example of a causally nonseparable bipartite process is the \textit{quantum switch}, which coherently superposes two fixed causal orders. A quantum control system determines the order in which a target system undergoes two operations $A$ then $B$. The possibility to have the control system in a quantum superposition indeed makes the resulting process causally nonseparable. An observation following the results in~\cite{VanderLugt2023DIQS} is that even if all the input/output systems $A_{I/O}$ and $B_{I/O}$ of the two operations $A$ and $B$ are dephased, then the resulting process remains causally indefinite. Indeed, the authors of~\cite{VanderLugt2023DIQS} consider a scenario in which $A$ and $B$ are measure-and-prepare operations in a fixed-basis, which operationally amounts to a full dephasing of the two slots of the switch. Despite this, the resulting correlations still violate an appropriate inequality, thereby certifying causal indefiniteness~\cite{Ebler_ftn}. Following up on this observation, we investigate which slots exactly can be dephased in general bipartite processes, and find that
\begin{figure}[t]
    \centering
    \includegraphics[width=.75\linewidth]{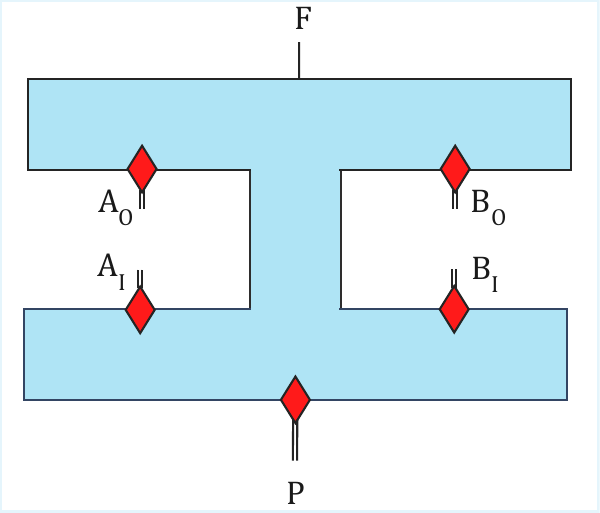}
     \caption{Any bipartite process matrix in which the global past $P$ and both parties’ input and output systems are fully dephased becomes causally separable, even if the global future $F$ is left coherent. The red diamonds illustrate dephasing in a fixed basis, which makes their incoming and outgoing systems effectively classical (double-stroke wires).
     }
    \label{fig:alldephased}
\end{figure}
\begin{figure*}[t]
    \centering
    \includegraphics[width=\textwidth]{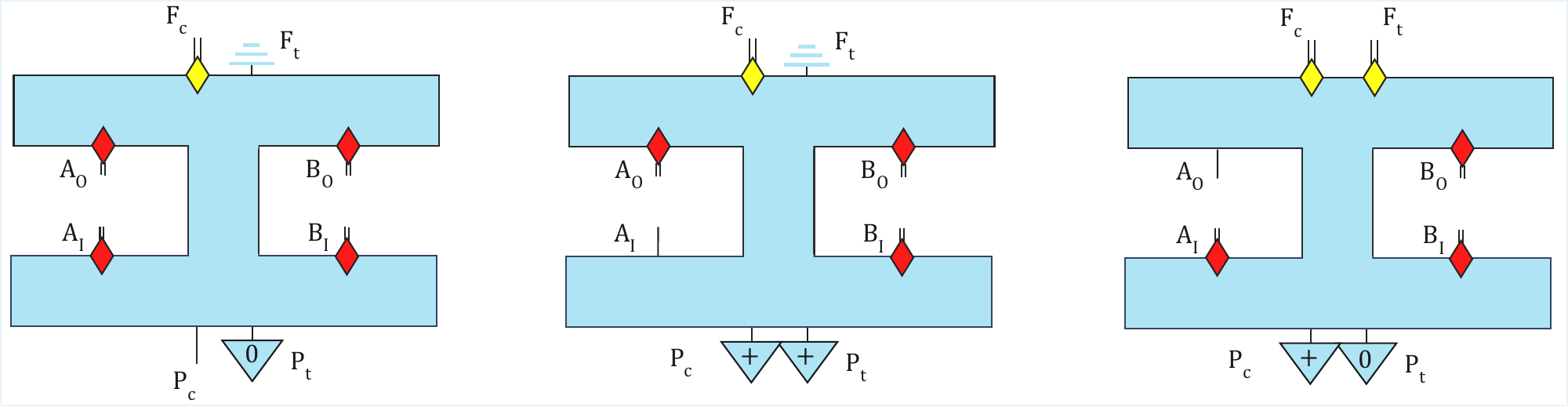}
    \caption{Three explicit constructions derived from a standard quantum switch showing that causal indefiniteness can survive when all but one systems are fully dephased -- as long as the single remaining coherent system is not $F$. \\ 
    (Left) Only the past control system remains coherent: the past is decomposed as $P = P_c \otimes P_t$, with the target state fixed to $\ket{0}^{P_t}$ while the control $P_c$ remains open; both internal parties' input and output systems are dephased in the computational ($Z$) basis (red diamonds), the future target $F_t$ is traced out, and the future control $F_c$ is dephased in the $X$ basis (yellow diamond). 
    (Middle) Only one party's input system $A^I$ remains coherent: the past state is fixed to $\ket{+,+}^{P_t P_c}$, the other parties' systems $A_O, B_I, B_O$ are dephased in the $Z$ basis, the future target is traced out, and the future control is dephased in the $X$ basis. 
    (Right) Only one party's output system $A_O$ remains coherent: the past state is fixed to $\ket{0,+}^{P_t P_c}$, the other parties' systems $A_I, B_I, B_O$ are dephased in the $Z$ basis, and both components of the future system are dephased in the $X$ basis. 
    In each case, all but one systems are effectively classical, yet causal nonseparability persists due to the coherence confined to the single untouched system. }
    \label{fig:examples}
\end{figure*}
 \begin{itemize}
     \item if all systems are dephased (in any fixed basis), or if only $F$ is left coherent, then any process matrix becomes causally separable (see Fig.~\ref{fig:alldephased});
     \item if all but one systems are dephased, and the single system that remains coherent is \emph{not} $F$, then there exists a process (which can be obtained from the quantum switch~\cite{Xbasis_ftn}) that remains causally nonseparable (see Fig.~\ref{fig:examples}).
 \end{itemize}

Going beyond the bipartite case, it is known that for $N\ge 3$, there exist classical causally nonseparable $N$-slot processes~\cite{Baumeler2014PRA,Baumeler_2016}, even without global past and future. Hence in such cases all systems can be dephased. One may however consider certain subclasses of processes, such as the physically well-motivated Quantum Circuits with Quantum Control of causal order (QC-QCs)~\cite{Wechs2021QCQC}, and ask the same question: how many systems can be dephased before any $N$-partite QC-QC becomes causally separable? It turns out that our previous bipartite result generalizes: if all systems are dephased, or if only $F$ is left coherent, then any $N$-partite QC-QC becomes causally separable. 
(And again if all but one system other than $F$ are dephased, then causal indefiniteness can still survive, as the previous bipartite examples can trivially be considered to be $N$-partite, by introducing classical phantomatic parties that do not take part to the processes.)

\medskip

Overall our results show that causal nonseparability can be quite robust to dephasing, i.e.\ to decohering certain systems. The general $N$-partite result in particular implies that one can construct causally nonseparable QC-QCs which only act on classical operations (with only classical open slots) -- which one may call Classical Circuits with Quantum Control of causal order (CC-QCs) -- as long as there remains a coherent open past.

\onecolumngrid
\vfill
\paragraph*{Acknowledgments.}
We thank Alastair A. Abbott and Hippolyte Dourdent for fruitful discussions.
We acknowledge funding from the Agence Nationale de la Recherche (project ANR-22-CE47-0012) and the PEPR integrated project EPiQ (grant number ANR-22-PETQ-0007) as part of Plan France 2030.

\medskip

For the purpose of open access, the authors have applied a CC-BY public copyright license to any Author Accepted Manuscript (AAM) version arising from this submission.


\clearpage
\onecolumngrid

\section*{Technical Material}

\bigskip

\section{Preliminaries}

\subsection{Notations}

${\cal L}({\cal H}^X)$ denotes the space of linear operators on a given Hilbert space ${\cal H}^X$, referred to as ``system $X$''. We only consider finite-dimensional Hilbert spaces, and denote by $d_X$ the dimension of system $X$ (i.e.\ of the Hilbert space ${\cal H}^X$). For two Hilbert spaces ${\cal H}^X$ and ${\cal H}^Y$, we simply denote by ${\cal H}^{XY}$ their tensor product ${\cal H}^X\otimes{\cal H}^Y$ (whose order is irrelevant, as long as we keep track of which subsystem lives in which space). We use the shorthand notations ${\cal H}^{A_{IO}^k} \coloneqq {\cal H}^{A_I^kA_O^k}$ and, for any subset ${\cal K}$ of ${\cal N}$, ${\cal H}^{A_{IO}^{\cal K}} \coloneqq \bigotimes_{k\in{\cal K}} {\cal H}^{A_{IO}^k}$. $\id^X$ denotes the identity operator in ${\cal H}^X$, $\Tr_X$ denotes the partial trace over system $X$, $\Tr$ with no subscript denotes the full trace, ${}^T$ denotes transposition.

In general, superscripts on operators indicate the Hilbert spaces in which these act (e.g.\ we may write $W^{PA_{IO}^{\cal N}F}$ to indicate that $W\in{\cal L}({\cal H}^{PA_{IO}^{\cal N}F})$), although these may be dropped when this is clear from the context.

\subsection{The Process Matrix Framework}

The process matrix framework~\cite{OCB2012} provides the most general description of higher-order quantum transformations -- or ``supermaps''~\cite{Chiribella2008supermap}, i.e.\ maps that take local quantum operations as inputs and return new quantum operations (or probabilities) -- without presupposing a fixed causal order.

We consider $N$ quantum operations ${\cal A}_k: {\cal L}({\cal H}^{A_I^k})\to{\cal L}({\cal H}^{A_O^k})$, for $k\in{\cal N}\coloneqq \{1,\ldots,N\}$, from some input Hilbert spaces ${\cal H}^{A_I^k}$ to some output Hilbert spaces ${\cal H}^{A_O^k}$. A supermap ${\cal W}$ transforms these into a new quantum operation ${\cal M} = {\cal W}({\cal A}_1,\ldots,{\cal A}_N): {\cal L}({\cal H}^P)\to{\cal L}({\cal H}^F)$ from some ``global past'' Hilbert space ${\cal H}^P$ to some ``global future'' Hilbert space ${\cal H}^F$. The process matrix $W$ describing this supermap is an operator in ${\cal L}({\cal H}^{PA_{IO}^{\cal N}F})$ that relates the Choi representations~\cite{Choi1975} $A_k$ of the input operations ${\cal A}_k\in{\cal L}({\cal H}^{A_{IO}^k})$ to the Choi representation $M\in{\cal L}({\cal H}^{PF})$ of the output operation ${\cal M}$, through~\cite{Araujo2017purification}
\begin{align}
    M = \Tr_{A_{IO}^{\cal N}}\left[(A_1\otimes\cdots\otimes A_N)^T\otimes\id^{PF}\cdot W\right].
\end{align}
We will talk indifferently about the operations and their Choi representation, and about the supermap ${\cal W}$ or its process matrix representation $W$ (which we may also simply call a process).

A process matrix is required to transform completely positive (CP) maps into a CP map, and trace-preserving (TP) maps into a TP map -- even when the input operations are in fact extended to other systems, and the process only acts locally on part of the operations.%
\footnote{Note that in the original formulation~\cite{OCB2012}, there were no global past and future spaces and process matrices were returning scalar values; they were required to return values between 0 and 1 (valid probabilities) when acting on CP, trace-non-increasing input operations, and the value 1 when acting on CPTP operations. That original formulation is recovered here by taking $P$ and $F$ to be trivial (1-dimensional) systems.}
This imposes some constraints on process matrices: a process matrix $W$ must be positive semidefinite (PSD) and must satisfy certain affine-linear constraints. In the case of 2 input operations -- the ``bipartite case'', where we rename the operations $A_1,A_2$ as $A,B$ -- these affine-linear constraints write~\cite{OCB2012,Araujo_2015,Araujo2017purification,Wechs_2019}
\begin{gather}
{}_{[1-A_O][1-B_O]}\Tr_F W = 0, \label{eq:cstr_valid_2} \\
{}_{[1-A_O]}\Tr_{B_{IO}F} W = 0, \quad
{}_{[1-B_O]}\Tr_{A_{IO}F} W = 0, \label{eq:cstr_valid_1} \\
{}_{[1-P]}\Tr_{A_{IO}B_{IO}F} W = 0, \label{eq:cstr_valid_0} \\
\Tr W = d_P\,d_{A_O}\,d_{B_O}, \label{eq:cstr_valid_normalisation}
\end{gather}
where we used the ``trace-and-replace'' notation:
\begin{align}
{}_X W \coloneqq (\Tr_X W) \otimes \frac{\mathbbm 1^X}{d_X},
\qquad {}_{[1-X]}W \coloneqq W - {}_X W .
\end{align}
In the general case of $N$ input operations -- the ``$N$-partite'' case -- the affine-linear validity conditions write
\begin{gather}
\forall\,{\cal K}\subseteq{\cal N}, {\cal K}\neq\emptyset, \ {}_{\prod_{k\in{\cal K}}[1-A_O^k]}\Tr_{A_{IO}^{{\cal N}\backslash{\cal K}}F} W = 0, \label{eq:cstr_validN_n} \\
{}_{[1-P]}\Tr_{A_{IO}^{\cal N}F} W = 0, \label{eq:cstr_validN_0} \\
\Tr W = d_P\,\prod_{k\in{\cal N}}d_{A_O^k}. \label{eq:cstr_validN_normalisation}
\end{gather}

\medskip

As mentioned above, the process matrix framework does not presuppose that the input operations are combined in any particular causal order. One can in fact identify different classes of processes, whether these are compatible with a fixed order between the input operations, compatible with a definite (possibly probabilistic and dynamical) causal order -- in which case the processes are said to be \emph{causally separable} -- or incompatible with any definite causal order -- so-called \emph{causally nonseparable processes}.

Causally separable processes can typically be understood as quantum circuits with classical control of causal order (QC-CCs)~\cite{OreshkovGiarmatzi_2016,Wechs2021QCQC}.%
\footnote{For $N\le 3$, the class of QC-CCs coincides precisely with that of causally separable processes. For $N\ge 4$, clearly QC-CCs are causally separable; however, whether all causally separable processes are QC-CCs remains an open question~\cite{Wechs_2019,Wechs2021QCQC}.}
Among causally nonseparable processes, some can be understood as quantum circuits with quantum control of causal order (QC-QCs)~\cite{Wechs2021QCQC}. More general processes also exist, whose physical interpretation remains unclear~\cite{OCB2012}.

In this work we focus on the two classes of QC-QCs and QC-CCs. As shown in~\cite{Wechs2021QCQC}, these can be characterised in terms of linear and positive-semidefinite constraints, which we detail below.

\subsection{Quantum circuits with quantum control of causal order}

\subsubsection{Bipartite case}

A matrix $W\in{\cal L}({\cal H}^{PA_{IO}B_{IO}F})$ is the process matrix of a bipartite QC-QC (with $N=2$ input operations $A,B$) if and only if $W$ is PSD and there exist PSD matrices $W_{(A,B)}\in{\cal L}({\cal H}^{PA_{IO}B_I})$, $W_{(B,A)}\in{\cal L}({\cal H}^{PB_{IO}A_I})$, $W_{(A)}\in{\cal L}({\cal H}^{PA_I})$, $W_{(B)}\in{\cal L}({\cal H}^{PB_I})$ such that
\begin{gather}
    \Tr_F W = W_{(A,B)}\otimes\id^{B_O} + W_{(B,A)}\otimes\id^{A_O}, \label{eq:cstr_QCQC2_2} \\
    \Tr_{B_I} W_{(A,B)} = W_{(A)}\otimes\id^{A_O}, \quad \Tr_{A_I} W_{(B,A)} = W_{(B)}\otimes\id^{B_O}, \label{eq:cstr_QCQC2_1} \\
    \Tr_{A_I} W_{(A)} + \Tr_{B_I} W_{(B)} = \id^P. \label{eq:cstr_QCQC2_0}
\end{gather}

\medskip

The so-called ``quantum switch''~\cite{Chiribella2013}, which we present in more details in Sec.~\ref{subsec:QS}, is a paradigmatic example of a bipartite QC-QC.

\subsubsection{$N$-partite case}

A matrix $W\in{\cal L}({\cal H}^{PA_{IO}^{\cal N}F})$ is the process matrix of an $N$-partite QC-QC if and only if $W$ is PSD and there exist PSD matrices $W_{({\cal K}_n,k_{n+1})}\in{\cal L}({\cal H}^{PA_{IO}^{{\cal K}_n}A_I^{k_{n+1}}})$, for all strict subsets%
\footnote{The notation ${\cal K}_n$ implicitly assumes that $|{\cal K}_n|=n$.}
${\cal K}_n$ of ${\cal N}$ and all $k_{n+1}\in{\cal N}\backslash{\cal K}_n$,  such that
\begin{gather}
    \Tr_F W = \sum_{k_N \in \mathcal{N}} W_{(\mathcal{N}\setminus\{k_N\},k_N)} \otimes \id^{A_O^{k_N}}, \label{eq:cstr_QCQCN_N} \\
    \forall\, \emptyset \subsetneq \mathcal{K}_n \subsetneq \mathcal{N}, \quad
\sum_{k_{n+1}\in \mathcal{N}\setminus \mathcal{K}_n} \Tr_{A_I^{k_{n+1}}} W_{(\mathcal{K}_n,k_{n+1})} = \sum_{k_n\in \mathcal{K}_n} W_{(\mathcal{K}_n\setminus\{k_n\},k_n)}\otimes\id^{A_O^{k_n}}, \label{eq:cstr_QCQCN_n} \\
    \sum_{k_1 \in \mathcal{N}} \Tr_{A_I^{k_1}} W_{(\emptyset,k_1)} = \id^P. \label{eq:cstr_QCQCN_0}
\end{gather}

\subsection{Quantum circuits with classical control of causal order}

\subsubsection{Bipartite case}

A matrix $W\in{\cal L}({\cal H}^{PA_{IO}B_{IO}F})$ is the process matrix of a bipartite QC-CC if and only if $W$ is PSD and there exist PSD matrices $W_{(A,B,F)}\in{\cal L}({\cal H}^{PA_{IO}B_{IO}F})$, $W_{(B,A,F)}\in{\cal L}({\cal H}^{PA_{IO}B_{IO}F})$, $W_{(A,B)}\in{\cal L}({\cal H}^{PA_{IO}B_I})$, $W_{(B,A)}\in{\cal L}({\cal H}^{PB_{IO}A_I})$, $W_{(A)}\in{\cal L}({\cal H}^{PA_I})$, $W_{(B)}\in{\cal L}({\cal H}^{PB_I})$ such that
\begin{gather}
    W = W_{(A,B,F)} + W_{(B,A,F)}, \label{eq:cstr_QCCC2_3} \\
    \Tr_F W_{(A,B,F)} = W_{(A,B)}\otimes\id^{B_O}, \quad \Tr_F W_{(B,A,F)} = W_{(B,A)}\otimes\id^{A_O}, \label{eq:cstr_QCCC2_2} \\
    \Tr_{B_I} W_{(A,B)} = W_{(A)}\otimes\id^{A_O}, \quad \Tr_{A_I} W_{(B,A)} = W_{(B)}\otimes\id^{B_O}, \label{eq:cstr_QCCC2_1} \\
    \Tr_{A_I} W_{(A)} + \Tr_{B_I} W_{(B)} = \id^P. \label{eq:cstr_QCCC2_0}
\end{gather}

\medskip

Note that this characterisation differs from that of bipartite QC-QCs, Eqs.~\eqref{eq:cstr_QCQC2_2}--\eqref{eq:cstr_QCQC2_0}, only by the first two lines. Compared to a QC-QC, which only decomposes into 2 PSD matrices $W_{(A,B)}$ and $W_{(B,A)}$ \emph{after tracing out $F$}, a QC-CC process matrix is required to also decompose into 2 PSD matrices $W_{(A,B,F)}$ and $W_{(B,A,F)}$ \emph{before tracing out $F$}; these must individually give $W_{(A,B)}$ and $W_{(B,A)}$ (together with some identity operator) after tracing out $F$, as in E.~\eqref{eq:cstr_QCCC2_2} above.

Notice also that if there is no $F$ (or if $F$ is trivial, 1-dimensional) then the distinction between bipartite QC-QCs and QC-CCs vanishes: any bipartite QC-QC with no $F$ is in fact a QC-CC.

\subsubsection{$N$-partite case}

A matrix $W\in{\cal L}({\cal H}^{PA_{IO}^{\cal N}F})$ is the process matrix of an $N$-partite QC-CC if and only if $W$ is PSD and there exist PSD matrices $W_{(k_1,\ldots,k_N,F)}\in{\cal L}({\cal H}^{PA_{IO}^{{\cal N}}F})$ and $W_{(k_1,\ldots,k_{n-1},k_n)}\in{\cal L}({\cal H}^{PA_{IO}^{\{k_1,\ldots,k_{n-1}\}}A_I^{k_n}})$, for all $n=1,\ldots,N$ and all sequences%
\footnote{The notation $(k_1,\ldots,k_n)$ implicitly assumes that all $k_i$'s in the (ordered) sequence are different.}
$(k_1,\ldots,k_n)$ of elements in ${\cal N}$,  such that
\begin{gather}
    W = \sum_{(k_1,\ldots,k_N)} W_{(k_1,\ldots,k_N,F)} \label{eq:cstr_QCCCN_Np1} \\
    \forall\,(k_1,\ldots,k_N), \quad \Tr_F W_{(k_1,\ldots,k_N,F)} = W_{(k_1,\ldots,k_N)} \otimes \id^{A_O^{k_N}}, \label{eq:cstr_QCCCN_N} \\
    \forall\, n = 1,\ldots,N-1, \ \forall\,(k_1,\ldots,k_n), \quad
\sum_{k_{n+1}\in \mathcal{N}\setminus \{k_1,\ldots,k_n\}} \Tr_{A_I^{k_{n+1}}} W_{(k_1,\ldots,k_n,k_{n+1})} = W_{(k_1,\ldots,k_n)}\otimes\id^{A_O^{k_n}}, \label{eq:cstr_QCCCN_n} \\
    \sum_{k_1 \in \mathcal{N}} \Tr_{A_I^{k_1}} W_{(k_1)} = \id^P. \label{eq:cstr_QCCCN_0}
\end{gather}

\subsection{Dephasing}
\label{subsec:dephasing}

In this work we consider the action of dephasing quantum systems -- making them effectively classical -- and investigate how this affects the potential causal nonseparability of quantum processes.

Dephasing is defined with respect to a specific basis. For a given system $S$ and a basis $\{\ket{i}\}_i$ of ${\cal H}^S$, we define the dephasing map $\Delta_S$ that acts on any $W\in{\cal L}({\cal H}^{ST})$, for any complementary space ${\cal H}^T$, as
\begin{align}
    \Delta_S(W) \coloneqq \sum_i \left(\ketbra{i}{i}^S\otimes\id^T\right) W \left(\ketbra{i}{i}^S\otimes\id^T\right) = \sum_i \,[\![i]\!]^S \otimes W_i^T, \label{eq:def_deph}
\end{align}
where we introduced the notation $[\![i]\!]^S\coloneq \ketbra{i}{i}$, and with $W_i^T = \big(\bra{i}^S\otimes\id^T\big) W \big(\ket{i}^S\otimes\id^T\big)$.

We note that the dephasing map is an idempotent CPTP map which is unital (it preserves the identity operator) and commutes with the partial trace, in the sense that for any systems $S,T,R$ and any $W\in{\cal L}({\cal H}^{STR})$,
\begin{align}
    \Tr_{T} [\Delta_S(W)] = \Delta_{S\backslash T}(\Tr_T W),
\end{align}
where we allow the systems $S$ and $T$ to be composed of subsystems $S_j$, $T_k$, and $S\backslash T$ denotes the set of $S_j$'s that are not among the $T_k$'s (as the latter are traced out before dephasing, in the right-hand-side above). This implies in particular%
\footnote{This can be seen by looking at the general validity constraints~\eqref{eq:cstr_validN_n}--\eqref{eq:cstr_validN_normalisation}, the QC-QC constraints~\eqref{eq:cstr_QCQCN_N}--\eqref{eq:cstr_QCQCN_0} or the QC-CC constraints~\eqref{eq:cstr_QCCCN_Np1}--\eqref{eq:cstr_QCCCN_0}, respectively, which all simply involve taking partial traces and tensoring with identity operators.}
that if $W$ is a valid process matrix, then so is $\Delta_S(W)$, for any system $S$; if $W$ is a QC-QC, then so is $\Delta_S(W)$; and if $W$ is a QC-CC, then so is $\Delta_S(W)$. In the latter two cases, if $W$ has a QC-QC or QC-CC decomposition in terms of matrices $W_{({\cal K}_n,k_{n+1})}$ or $W_{(k_1,\ldots,k_n)}/W_{(k_1,\ldots,k_N,F)}$, then $\Delta_S(W)$ has a similar QC-QC or QC-CC decomposition in terms of matrices $\Delta_S(W_{({\cal K}_n,k_{n+1})})$ or $\Delta_S(W_{(k_1,\ldots,k_n)})/\Delta_S(W_{(k_1,\ldots,k_N,F)})$, respectively.

\medskip

In the proofs below, rather than involving the dephasing map $\Delta_S$ explicitly, we will consider process matrices $W$ with ``already dephased'' system(s) $S$. By this, we mean that they are of the form
\begin{align}
    W = \Delta_S(W) = \sum_i \,[\![i]\!]^S \otimes W_i^T,
\end{align}
as in Eq.~\eqref{eq:def_deph}: we say that they are diagonal in system $S$, in the basis $\{\ket{i}\}_i$. According to the previous remark, if a QC-QC or a QC-CC is diagonal in some given system and some given basis, then one can take all matrices $W_{({\cal K}_n,k_{n+1})}$ or $W_{(k_1,\ldots,k_n)}/W_{(k_1,\ldots,k_N,F)}$ in its QC-QC or QC-CC decomposition to also be diagonal in the same system, in the same basis.

\section{Proof that any bipartite process with dephased $P, A_I, B_I$ is a QC-QC}

In this appendix we prove that any bipartite process matrix $W\in{\cal L}({\cal H}^{PA_{IO}B_{IO}F})$ with dephased systems $P, A_I$ and $B_I$ is a QC-QC.

Note that in the case with no future space ${\cal H}^F$, this implies that any such $W$ matrix is a QC-CC -- hence, is causally separable. If, further, there is no past space ${\cal H}^P$, then we get that any process matrix $W\in{\cal L}({\cal H}^{A_{IO}B_{IO}})$ with dephased input systems $A_I$ and $B_I$ is a convex mixture of fixed-order processes: in that case we recover the result of Baumann and Brukner~\cite{Baumann2016Appearance}, which was a strengthening of Oreshkov \emph{et al.}'s original result that classical (i.e., fully dephased) bipartite processes are causally separable~\cite{OCB2012}.

As a warm-up and in order to provide some intuition, we start by proving our claim above in the simpler case where all systems (possibly except for $F$%
\footnote{The proofs below do not require one to assume that the future system $F$ is dephased; of course the results also hold, \emph{a fortiori}, if $F$ is also dephased.}%
) are dephased.

\subsection{All systems (except for $F$) being dephased}
\label{subsec:N2_all_dephased}

Consider a bipartite process matrix $W$. The starting point is to note that the validity constraint ${}_{[1-A_O][1-B_O]}\Tr_F W=0$ of Eq.~\eqref{eq:cstr_valid_2} implies that $\Tr_F W$ can be decomposed (in a non-unique manner) as%
\footnote{One way to see this is by noting that if one decomposes $W$ onto a Hilbert-Schmidt basis for each subsystem, then Eq.~\eqref{eq:cstr_valid_2} requires all terms in such a decomposition to have an identity operator on either $A_O$ or $B_O$ (or on both)~\cite{OCB2012}.}
\begin{align}
    \Tr_F W = S^{PA_IB_IA_O}\otimes \id^{B_O} + T^{PA_IB_IB_O}\otimes \id^{A_O},
\end{align}
for some (not necessarily PSD) matrices $S,T$. Furthermore, the validity constraints ${}_{[1-A_O]}\Tr_{B_{IO}F} W=0$ and ${}_{[1-B_O]}\Tr_{A_{IO}F} W=0$ of Eq.~\eqref{eq:cstr_valid_1} imply that $S$ and $T$ satisfy ${}_{[1-A_O]}\Tr_{B_I} S=0$ and ${}_{[1-B_O]}\Tr_{A_I} T=0$, respectively, so that $\Tr_{B_I} S=\frac{1}{d^{A_O}}\Tr_{A_OB_I} S \otimes \id^{A_O}$ and $\Tr_{A_I} T=\frac{1}{d^{B_O}}\Tr_{B_OA_I} T \otimes \id^{B_O}$. Also, the last validity constraint ${}_{[1-P]}\Tr_{A_{IO}B_{IO}F}W = 0$ of Eq.~\eqref{eq:cstr_valid_0} together with the normalisation constraint $\Tr W = d^Pd^{A_O}d^{B_O}$ imply that $\frac{1}{d^{A_O}d^{B_O}} \Tr_{A_{IO}B_{IO}} \left(S\otimes \id^{B_O} + T\otimes \id^{A_O}\right) = \id^P$.

Assume that systems $P, A_I, B_I, A_O, B_O$ are dephased in the bases $\{\ket{p}^P\}_p$, $\{\ket{i}^{A_I}\}_i$, $\{\ket{j}^{B_I}\}_j$, $\{\ket{k}^{A_O}\}_k$, $\{\ket{l}^{B_O}\}_l$, respectively, so that $\Tr_F W$ is diagonal in these bases. $S$ and $T$ can then also be taken to be diagonal in the same bases, so that one can write
\begin{align}
    S^{PA_IB_IA_O} = \sum_{p,i,j,k} s_{p,i,j,k} \, [\![p,i,j,k]\!]^{PA_IB_IA_O}, \quad T^{PA_IB_IB_O} = \sum_{p,i,j,l} t_{p,i,j,l} \, [\![p,i,j,l]\!]^{PA_IB_IB_O},
\end{align}
for some real coefficients $s_{p,i,j,k}$ and $t_{p,i,j,l}$, such that 
\begin{align}
    \Tr_F W = \sum_{p,i,j,k,l} (s_{p,i,j,k} + t_{p,i,j,l}) \, [\![p,i,j,k,l]\!]^{PA_IB_IA_OB_O}.
\end{align}
While the coefficients $s_{p,i,j,k}$ and $t_{p,i,j,l}$ may be negative, their sums $s_{p,i,j,k} + t_{p,i,j,l}$ must be nonnegative (as required by $\Tr_F W\ge0$). As it turns out, one can always find other coefficients $s'_{p,i,j,k}$ and $t'_{p,i,j,l}$ that are nonnegative and such that $s'_{p,i,j,k} + t'_{p,i,j,l} = s_{p,i,j,k} + t_{p,i,j,l}$ for all $p,i,j,k,l$.
A possible choice is indeed to introduce, for all $p,i,j$, $\underline{s}_{p,i,j} \coloneqq \min_k s_{p,i,j,k}$, and then define $s'_{p,i,j,k}\coloneqq s_{p,i,j,k} - \underline{s}_{p,i,j}$ and $t'_{p,i,j,l}\coloneqq \underline{s}_{p,i,j} + t_{p,i,j,l}$.%
\footnote{One indeed clearly has $s'_{p,i,j,k}\ge0$, $s'_{p,i,j,k} + t'_{p,i,j,l} = s_{p,i,j,k} + t_{p,i,j,l}$; and by denoting $k^*$ the index such that $\underline{s}_{p,i,j} = s_{p,i,j,k^*}$, $t'_{p,i,j,l} = s_{p,i,j,k^*} + t_{p,i,j,l} \ge0$.}

With this, let us then define
\begin{align}
    W_{(A,B)} \coloneqq \sum_{p,i,j,k} s'_{p,i,j,k} \, [\![p,i,j,k]\!]^{PA_IB_IA_O} \ge0, \qquad W_{(B,A)} = \sum_{p,i,j,l} t'_{p,i,j,l} \, [\![p,i,j,l]\!]^{PA_IB_IB_O} \ge0,
\end{align}
such that 
\begin{align}
    W_{(A,B)} \otimes \id^{B_O} + W_{(B,A)} \otimes \id^{A_O} & = \sum_{p,i,j,k,l} (s'_{p,i,j,k} + t'_{p,i,j,l}) \, [\![p,i,j,k,l]\!]^{PA_IB_IA_OB_O} \notag \\
     & = \sum_{p,i,j,k,l} (s_{p,i,j,k} + t_{p,i,j,l}) \, [\![p,i,j,k,l]\!]^{PA_IB_IA_OB_O} = \Tr_F W.
\end{align}
Furthermore,
\begin{align}
    \Tr_{B_I} W_{(A,B)} & = \Tr_{B_I} \Big(S - \sum_{p,i,j,k} \underline{s}_{p,i,j} \, [\![p,i,j,k]\!]^{PA_IB_IA_O} \Big) = W_{(A)} \otimes \id^{A_O}, \notag \\
    \Tr_{A_I} W_{(B,A)} & = \Tr_{A_I} \Big(T + \sum_{p,i,j,l} \underline{s}_{p,i,j} \, [\![p,i,j,l]\!]^{PA_IB_IB_O} \Big) = W_{(B)} \otimes \id^{B_O} \label{eq:TrBI_W_AB_proofSecII}
\end{align}
with
\begin{align}
    W_{(A)} \coloneqq \frac{1}{d^{A_O}}\Tr_{A_OB_I} S - \sum_{p,i,j} \underline{s}_{p,i,j} \, [\![p,i]\!]^{PA_I} \ge 0, \qquad
    W_{(B)} \coloneqq \frac{1}{d^{B_O}}\Tr_{B_OA_I} T + \sum_{p,i,j} \underline{s}_{p,i,j} \, [\![p,j]\!]^{PB_I} \ge 0, \label{eq:def_WA_WB_proofSecII}
\end{align}
such that
\begin{align}
    \Tr_{A_I} W_{(A)} + \Tr_{B_I} W_{(B)} & = \frac{1}{d^{A_O}}\Tr_{A_IA_OB_I} S + \frac{1}{d^{B_O}}\Tr_{B_IB_OA_I} T \notag \\
    & = \frac{1}{d^{A_O}d^{B_O}} \Tr_{A_{IO}B_{IO}} \left(S\otimes \id^{B_O} + T\otimes \id^{A_O}\right) = \id^P. \label{eq:TrAI_WA_TrBI_WB_proofSecII}
\end{align}

All in all, we thus obtain a decomposition of $W$ in terms of PSD matrices $W_{(A,B)}, W_{(B,A)}, W_{(A)}$ and $W_{(B)}$ satisfying Eqs.~\eqref{eq:cstr_QCQC2_2}--\eqref{eq:cstr_QCQC2_0}, which shows that $W$ is a QC-QC.

\subsection{With only systems $P,A_I,B_I$ being dephased}

When only the systems $P,A_I,B_I$ are dephased, we can proceed in a similar way as above, except that we start with matrices $S$ and $T$ that can be decomposed as
\begin{align}
    S^{PA_IB_IA_O} = \sum_{p,i,j} \, [\![p,i,j]\!]^{PA_IB_I} \otimes S_{p,i,j}^{A_O}, \quad T^{PA_IB_IB_O} = \sum_{p,i,j} \, [\![p,i,j]\!]^{PA_IB_I} \otimes T_{p,i,j}^{B_O},
\end{align}
for some (not necessarily PSD) matrices $S_{p,i,j}^{A_O}$ and $T_{p,i,j}^{B_O}$, such that 
\begin{align}
    \Tr_F W = \sum_{p,i,j} \, [\![p,i,j]\!]^{PA_IB_I} \otimes (S_{p,i,j}^{A_O}\otimes \id^{B_O} + T_{p,i,j}^{B_O}\otimes \id^{A_O}).
\end{align}
One can then proceed with the same arguments as above, after diagonalising each matrix $S_{p,i,j}$ and $T_{p,i,j}$ independently (so that $S_{p,i,j}^{A_O}\otimes \id^{B_O}$ and $T_{p,i,j}^{B_O}\otimes \id^{A_O}$ above are diagonalised in the same basis -- a generally different basis for each value of $p,i,j$).

Alternatively (and more explicitly), one can note that while the matrices $S_{p,i,j}$ and $T_{p,i,j}$ may be non-PSD, their combination $S_{p,i,j}^{A_O}\otimes \id^{B_O} + T_{p,i,j}^{B_O}\otimes \id^{A_O}$ must be PSD. In a similar way as above, one can always find other matrices $S_{p,i,j}^{\prime\,A_O}$ and $T_{p,i,j}^{\prime\,B_O}$ that are PSD and such that $S_{p,i,j}^{\prime\,A_O}\otimes \id^{B_O} + T_{p,i,j}^{\prime\,B_O}\otimes \id^{A_O} = S_{p,i,j}^{A_O}\otimes \id^{B_O} + T_{p,i,j}^{B_O}\otimes \id^{A_O}$ for all $p,i,j$.
A possible choice here is to consider, for all $p,i,j$, the minimum eigenvalue $\underline{s}_{p,i,j}$ of $S_{p,i,j}$, and then define $S_{p,i,j}^{\prime\,A_O}\coloneqq S_{p,i,j}^{A_O} - \underline{s}_{p,i,j}\id^{A_O}$ and $T_{p,i,j}^{\prime\,B_O}\coloneqq \underline{s}_{p,i,j}\id^{B_O} + T_{p,i,j}^{B_O}$.%
\footnote{One indeed clearly has $S_{p,i,j}^{\prime\,A_O}\ge0$, $S_{p,i,j}^{\prime\,A_O}\otimes \id^{B_O} + T_{p,i,j}^{\prime\,B_O}\otimes \id^{A_O} = S_{p,i,j}^{A_O}\otimes \id^{B_O} + T_{p,i,j}^{B_O}\otimes \id^{A_O}$; and denoting by $\ket{\psi^*}^{A_O}$ the eigenstate of $S_{p,i,j}^{A_O}$ with eigenvalue $\underline{s}_{p,i,j}$, then for any state $\ket{\varphi}^{B_O}$, $\bra{\varphi}T_{p,i,j}^{\prime\,B_O}\ket{\varphi} = \underline{s}_{p,i,j} + \bra{\varphi}T_{p,i,j}^{B_O}\ket{\varphi} = \bra{\psi^*}S_{p,i,j}^{A_O}\ket{\psi^*} + \bra{\varphi}T_{p,i,j}^{B_O}\ket{\varphi} = \bra{\psi^*,\varphi}(S_{p,i,j}^{A_O}\otimes \id^{B_O} + T_{p,i,j}^{B_O}\otimes \id^{A_O})\ket{\psi^*,\varphi} \ge0$, so that $T_{p,i,j}^{\prime\,B_O}\ge 0$.}

Define now
\begin{align}
W_{(A,B)}
&\coloneqq
\sum_{p,i,j} \, [\![p,i,j]\!]^{P A_I B_I}\otimes S_{p,i,j}^{\prime\,A_O}
\;\ge\;0,
&
W_{(B,A)}
&\coloneqq
\sum_{p,i,j} \, [\![p,i,j]\!]^{P A_I B_I}\otimes T_{p,i,j}^{\prime\,B_O}
\;\ge\;0,
\label{eq:WAB_WBA_blocks}
\end{align}
such that 
\begin{align}
    W_{(A,B)} \otimes \id^{B_O} + W_{(B,A)} \otimes \id^{A_O} & = \sum_{p,i,j} \, [\![p,i,j]\!]^{PA_IB_I} \otimes (S_{p,i,j}^{\prime\,A_O}\otimes \id^{B_O} + T_{p,i,j}^{\prime\,B_O}\otimes \id^{A_O}) \notag \\
     & = \sum_{p,i,j} \, [\![p,i,j]\!]^{PA_IB_I} \otimes (S_{p,i,j}^{A_O}\otimes \id^{B_O} + T_{p,i,j}^{B_O}\otimes \id^{A_O}) = \Tr_F W.
\end{align}
Furthermore,
\begin{align}
    \Tr_{B_I} W_{(A,B)} & = \Tr_{B_I} \Big(S - \sum_{p,i,j} \underline{s}_{p,i,j} \, [\![p,i,j]\!]^{P A_I B_I}\otimes \id^{A_O} \Big) = W_{(A)} \otimes \id^{A_O}, \notag \\
    \Tr_{A_I} W_{(B,A)} & = \Tr_{A_I} \Big(T + \sum_{p,i,j} \underline{s}_{p,i,j} \, [\![p,i,j]\!]^{PA_IB_I}\otimes \id^{B_O} \Big) = W_{(B)} \otimes \id^{B_O},
\end{align}
as in Eq.~\eqref{eq:TrBI_W_AB_proofSecII} above, with $W_{(A)}$ and $W_{(B)}$ defined as in Eq.~\eqref{eq:def_WA_WB_proofSecII}, satisfying $\Tr_{A_I} W_{(A)} + \Tr_{B_I} W_{(B)} = \id^P$ as in Eq.~\eqref{eq:TrAI_WA_TrBI_WB_proofSecII}.
We thus again obtain a decomposition of $W$ in terms of PSD matrices $W_{(A,B)}, W_{(B,A)}, W_{(A)}$ and $W_{(B)}$ satisfying Eqs.~\eqref{eq:cstr_QCQC2_2}--\eqref{eq:cstr_QCQC2_0}, which shows that $W$ is a QC-QC.

\section{Proof that any fully-dephased (except for $F$) QC-QC is a QC-CC}

We prove here our claim that as soon as all systems except for $F$ (or including $F$, \emph{a fortiori}) are dephased, any QC-QC is in fact a QC-CC.

For simplicity we start with the bipartite case.
Note that in that case, by recalling the previous result that any fully dephased (except for $F$) bipartite process matrix is a QC-QC (see Sec~\ref{subsec:N2_all_dephased}), the claim here can in fact be strengthened: any bipartite fully dephased (except for $F$) bipartite process is a QC-CC.%
\footnote{Note that the same cannot hold for more parties, as it is known that there exist classical tripartite processes with indefinite causal order~\cite{Baumeler2014PRA,Baumeler_2016}.}

\subsection{Bipartite case}

Consider a bipartite QC-QC $W$, with a decomposition in terms of PSD matrices $W_{(A,B)}, W_{(B,A)}, W_{(A)}$ and $W_{(B)}$ satisfying Eqs.~\eqref{eq:cstr_QCQC2_2}--\eqref{eq:cstr_QCQC2_0} -- in particular, such that $\Tr_F W = W_{(A,B)}\otimes\id^{B_O} + W_{(B,A)}\otimes\id^{A_O}$.

Assuming that all systems (possibly except for $F$) are dephased, we can write
\begin{align}
    W & = \sum_{p,i,j,k,l} [\![p,i,j,k,l]\!]^{PA_IB_IA_OB_O} \otimes W_{p,i,j,k,l}^F \,, \notag \\
    \Tr_F W & = \sum_{p,i,j,k,l} w_{p,i,j,k,l} \, [\![p,i,j,k,l]\!]^{PA_IB_IA_OB_O} , \notag \\
    W_{(A,B)}\otimes \id^{B_O} & = \sum_{p,i,j,k,l} w_{(A,B);p,i,j,k}\, [\![p,i,j,k,l]\!]^{PA_IB_IA_OB_O} , \notag \\
    W_{(B,A)}\otimes \id^{A_O} & = \sum_{p,i,j,k,l} w_{(B,A);p,i,j,l}\, [\![p,i,j,k,l]\!]^{PA_IB_IA_OB_O}
\end{align}
for some PSD matrices $W_{p,i,j,k,l}^F$, with $w_{p,i,j,k,l} = \Tr \,W_{p,i,j,k,l}^F$, and for some nonnegative coefficients $w_{(A,B);p,i,j,k}$, $w_{(B,A);p,i,j,l}$ such that $w_{p,i,j,k,l} =  w_{(A,B);p,i,j,k} + w_{(B,A);p,i,j,l}$.
With this, let us define%
\footnote{If $w_{p,i,j,k,l} = 0$, then necessarily $w_{(A,B);p,i,j,k} = w_{(B,A);p,i,j,l} = 0$ as well, and we take the corresponding terms in $W_{(A,B,F)}$ and $W_{(B,A,F)}$ to also be 0. \label{ftn:zero_denom}}
\begin{align}
    W_{(A,B,F)} & \coloneqq \sum_{p,i,j,k,l} \frac{w_{(A,B);p,i,j,k}}{w_{p,i,j,k,l}} \, [\![p,i,j,k,l]\!]^{PA_IB_IA_OB_O} \otimes W_{p,i,j,k,l}^F\,, \notag \\
    W_{(B,A,F)} & \coloneqq \sum_{p,i,j,k,l} \frac{w_{(B,A);p,i,j,l}}{w_{p,i,j,k,l}} \, [\![p,i,j,k,l]\!]^{PA_IB_IA_OB_O} \otimes W_{p,i,j,k,l}^F\,.
\end{align}
We clearly have $W_{(A,B,F)}\ge0$, $W_{(B,A,F)}\ge0$, $W_{(A,B,F)} + W_{(B,A,F)} = W$, $\Tr_F W_{(A,B,F)} = W_{(A,B)}\otimes \id^{B_O}$ and $\Tr_F W_{(B,A,F)} = W_{(B,A)}\otimes \id^{A_O}$.

We thus obtain a decomposition for $W$ in terms of PSD matrices $W_{(A,B,F)}, W_{(B,A,F)}, W_{(A,B)}, W_{(B,A)}, W_{(A)}$ and $W_{(B)}$ satisfying Eqs.~\eqref{eq:cstr_QCCC2_3}--\eqref{eq:cstr_QCCC2_0}, which shows that $W$ is a QC-CC.

\subsection{$N$-partite case}

Consider an $N$-partite QC-QC $W$, with a decomposition in terms of PSD matrices $W_{({\cal K}_{n-1},k_n)}$ satisfying Eqs.~\eqref{eq:cstr_QCQCN_N}--\eqref{eq:cstr_QCQCN_0}, and assume that all systems except for $F$ are dephased (which implies that all these matrices $W_{({\cal K}_{n-1},k_n)}$ can be taken to be diagonal).

The proof proceeds by induction: we show that for any $n = 0,\ldots,N$, one can construct (diagonal) PSD matrices $W_{(k_1,\ldots,k_n,k_{n+1})}$ such that $W_{({\cal K}_n,k_{n+1})} = \sum_{(k_1,\ldots,k_n)\in{\cal K}_n} W_{(k_1,\ldots,k_n,k_{n+1})}$ and $\sum_{k_{n+1}} \Tr_{A_I^{k_{n+1}}} W_{(k_1,\ldots,k_n,k_{n+1})} = W_{(k_1,\ldots,k_n)}\otimes\id^{A_O^{k_n}}$ (or such that $\sum_{k_1} \Tr_{A_I^{k_1}} W_{(k_1)} = \id^P$ in the $n=0$ case, or $W = \sum_{(k_1,\ldots,k_N)} W_{(k_1,\ldots,k_N,F)}$ and $\Tr_F W_{(k_1,\ldots,k_N,F)} = W_{(k_1,\ldots,k_N)}\otimes\id^{A_O^{k_N}}$ in the $n=N$ case, with $W_{(k_1,\ldots,k_N,F)}\equiv W_{(k_1,\ldots,k_N,k_{N+1})}$).
These will provide a decomposition for $W$ that satisfies Eqs.~\eqref{eq:cstr_QCCCN_Np1}--\eqref{eq:cstr_QCCCN_0}, thereby proving that $W$ is indeed a QC-CC.

\medskip

\emph{Base cases:}

Clearly for $n=0$ and $n=1$ the induction hypothesis holds, as can be seen by defining $W_{(k_1)}\coloneqq W_{(\emptyset,k_1)}$ and $W_{(k_1,k_2)}\coloneqq W_{(\{k_1\},k_2)}$ and by referring to Eqs.~\eqref{eq:cstr_QCQCN_n}--\eqref{eq:cstr_QCQCN_0}.

For $n=2$ the induction hypothesis can be proven in a similar way as in the bipartite proof of the previous subsection (by mapping any $(k_1,k_2,k_3)$ to $(A,B,F)$ or $(B,A,F)$). We do not write it explicitly, as this is in fact encompassed in the general induction step below.

\medskip\pagebreak

\emph{Induction step:}

Suppose that the induction hypothesis holds for some value $n-1<N$, so that one can construct (diagonal) PSD matrices $W_{(k_1,\ldots,k_n)}$ satisfying the appropriate constraints -- in particular, $W_{({\cal K}_{n-1},k_n)} = \sum_{(k_1,\ldots,k_{n-1})\in{\cal K}_{n-1}} W_{(k_1,\ldots,k_{n-1},k_n)}$.

Let us denote by $\{\ket{p}^P\}$, $\{\ket{i_{A_I^k}}^{A_I^k}\}$, $\{\ket{i_{A_O^k}}^{A_O^k}\}$ the various bases in which each system is dephased, and define, for $\mathcal K_n=\{k_1,\ldots,k_n\}$ and a multi-index $\vec{i}_{\mathcal K_n} \coloneqq (p,i_{A_I^{k_1}},i_{A_O^{k_1}},\ldots,i_{A_I^{k_n}},i_{A_O^{k_n}})$, the basis projector
\begin{align}
    [\![\vec i_{\mathcal K_n}]\!]^{P A_{IO}^{\mathcal K_n}} \coloneqq [\![p]\!]^P \bigotimes_{t=1}^n
\big([\![i_{A_I^{k_t}}]\!]^{A_I^{k_t}}\otimes[\![i_{A_O^{k_t}}]\!]^{A_O^{k_t}}\big).
\end{align}
With this, we can write (for each ${\cal K}_n,k_{n+1}\notin{\cal K}_n$ and $(k_1,\ldots,k_n)\in{\cal K}_n$; and for the case $n=N$, replacing $W_{({\cal N},k_{N+1})}$ by $W$ and $A^{k_{N+1}}_I$ by $F$):
\begin{align}
    W_{({\cal K}_n,k_{n+1})} & = \sum_{\vec i_{{\cal K}_n}}\, [\![\vec i_{{\cal K}_n}]\!]^{PA_{IO}^{{\cal K}_n}} \otimes W_{({\cal K}_n,k_{n+1});\vec i_{{\cal K}_n}}^{A^{k_{n+1}}_I} \ , \notag \\
    \Tr_{A^{k_{n+1}}_I} W_{({\cal K}_n,k_{n+1})} & = \sum_{\vec i_{{\cal K}_n}}\, w_{({\cal K}_n,k_{n+1});\vec i_{{\cal K}_n}}\,[\![\vec i_{{\cal K}_n}]\!]^{PA_{IO}^{{\cal K}_n}} \ , \notag \\
    W_{(k_1,\ldots,k_n)} \otimes \id^{A^{k_n}_O} & = \sum_{\vec i_{{\cal K}_n}}\,w_{(k_1,\ldots,k_n);\vec i_{{\cal K}_n}}\, [\![\vec i_{{\cal K}_n}]\!]^{PA_{IO}^{{\cal K}_n}} \ ,
\end{align}
for some PSD matrices $W_{({\cal K}_n,k_{n+1});\vec i_{{\cal K}_n}}^{A^{k_{n+1}}_I}$, with $w_{({\cal K}_n,k_{n+1});\vec i_{{\cal K}_n}} = \Tr\, W_{({\cal K}_n,k_{n+1});\vec i_{{\cal K}_n}}^{A^{k_{n+1}}_I}$ and for some nonnegative coefficients $w_{(k_1,\ldots,k_n);\vec i_{{\cal K}_n}}$.
Eq.~\eqref{eq:cstr_QCQCN_n} together with the induction hypothesis that $W_{({\cal K}_{n-1},k_n)} = \sum_{(k_1,\ldots,k_{n-1})\in{\cal K}_{n-1}} W_{(k_1,\ldots,k_{n-1},k_n)}$ then imply%
\footnote{For the case $n=N$, the sums over $k_{N+1}$ below reduce to just one term, corresponding to $k_{N+1}\equiv F$.}
\begin{align}
    \forall\,\vec i_{{\cal K}_n}, \qquad \sum_{k_{n+1}} w_{({\cal K}_n,k_{n+1});\vec i_{{\cal K}_n}} = \sum_{(k_1,\ldots,k_n)} w_{(k_1,\ldots,k_n);\vec i_{{\cal K}_n}} \quad \eqqcolon w_{{\cal K}_n;\vec i_{{\cal K}_n}}.
\end{align}
Let us then define%
\footnote{As in Footnote~\ref{ftn:zero_denom}, if $w_{{\cal K}_n;\vec i_{{\cal K}_n}} = 0$, then we take the corresponding terms in $W_{(k_1,\ldots,k_n,k_{n+1})}$ to also be 0.}
\begin{align}
    W_{(k_1,\ldots,k_n,k_{n+1})} \coloneqq \sum_{\vec i_{{\cal K}_n}}\, \frac{w_{(k_1,\ldots,k_n);\vec i_{{\cal K}_n}}}{w_{{\cal K}_n;\vec i_{{\cal K}_n}}} \, [\![\vec i_{{\cal K}_n}]\!]^{PA_{IO}^{{\cal K}_n}} \otimes W_{({\cal K}_n,k_{n+1});\vec i_{{\cal K}_n}}^{A^{k_{n+1}}_I}  \ge 0\, .
\end{align}
These satisfy
\begin{align}
    \sum_{(k_1,\ldots,k_n)\in{\cal K}_n} W_{(k_1,\ldots,k_n,k_{n+1})} = \sum_{\vec i_{{\cal K}_n}}\, \underbrace{\frac{\sum_{(k_1,\ldots,k_n)} w_{(k_1,\ldots,k_n);\vec i_{{\cal K}_n}}}{w_{{\cal K}_n;\vec i_{{\cal K}_n}}}}_{1} \, [\![\vec i_{{\cal K}_n}]\!]^{PA_{IO}^{{\cal K}_n}} \otimes W_{({\cal K}_n,k_{n+1});\vec i_{{\cal K}_n}}^{A^{k_{n+1}}_I} = W_{({\cal K}_n,k_{n+1})}
\end{align}
and
\begin{align}
    \sum_{k_{n+1}} \Tr_{A^{k_{n+1}}_I} W_{(k_1,\ldots,k_n,k_{n+1})} = \sum_{\vec i_{{\cal K}_n}}\, \frac{w_{(k_1,\ldots,k_n);\vec i_{{\cal K}_n}}}{w_{{\cal K}_n;\vec i_{{\cal K}_n}}} \underbrace{\sum_{k_{n+1}} w_{({\cal K}_n,k_{n+1});\vec i_{{\cal K}_n}}}_{w_{{\cal K}_n;\vec i_{{\cal K}_n}}}\, [\![\vec i_{{\cal K}_n}]\!]^{PA_{IO}^{{\cal K}_n}} = W_{(k_1,\ldots,k_n)} \otimes \id^{A^{k_n}_O},
\end{align}
as required: this shows that the induction hypothesis holds for the value $n$ -- and hence by further induction, for any value $n = 0,\ldots,N$, which completes the proof.

\section{Examples of partially-dephased quantum switches \\ that remain causally nonseparable}

\subsection{The quantum switch}
\label{subsec:QS}

The quantum switch~\cite{Chiribella2013} is a canonical example of a bipartite QC-QC, in which the order between the two operations $A$ and $B$ applied to a target system is coherently controlled by the state of a control qubit: if that state is $\ket{0}$, then the process applies $A$ before $B$, while if that state is $\ket{1}$, the process applies $B$ before $A$. If the control qubit is prepared in a quantum superposition of $\ket{0}$ and $\ket{1}$, then one obtains a superposition of the two orders.

The quantum switch (with a qubit target system) can be described as the following process matrix~\cite{Araujo_2015,OreshkovGiarmatzi_2016,Wechs2021QCQC}:
\begin{align}
W_{\mathrm{QS}} =
\dketbroj{w_{\mathrm{QS}}} \quad \text{with} \quad
\dket{w_{\mathrm{QS}}}
=
\sum_{i,j,k=0,1}
\Big(
& \ket{0}^{P_c}
  \ket{i}^{P_t}
  \ket{i}^{A_I}
  \ket{j}^{A_O}
  \ket{j}^{B_I}
  \ket{k}^{B_O}
  \ket{k}^{F_t}
  \ket{0}^{F_c}
\nonumber\\[-3mm]
&+
  \ket{1}^{P_c}
  \ket{i}^{P_t}
  \ket{i}^{B_I}
  \ket{j}^{B_O}
  \ket{j}^{A_I}
  \ket{k}^{A_O}
  \ket{k}^{F_t}
  \ket{1}^{F_c}
\Big)
\end{align}
(with implicit tensor products), where the global past and future spaces are composed of both a control and a target component: ${\cal H}^P={\cal H}^{P_cP_t}$, ${\cal H}^F={\cal H}^{F_cF_t}$.
It can be verified that $W_{\mathrm{QS}}$ has a QC-QC decomposition as in Eqs.~\eqref{eq:cstr_QCQC2_2}--\eqref{eq:cstr_QCQC2_0}, with
\begin{align}
    W_{(A,B)} &= \sum_{i,j=0,1} [\![0]\!]^{P_c} [\![i]\!]^{P_t} [\![i]\!]^{A_I} [\![j]\!]^{A_O} [\![j]\!]^{B_I}, & W_{(B,A)} &= \sum_{i,j=0,1} [\![1]\!]^{P_c} [\![i]\!]^{P_t} [\![i]\!]^{B_I} [\![j]\!]^{B_O} [\![j]\!]^{A_I}, \notag \\
    W_{(A)} &= \sum_{i=0,1} [\![0]\!]^{P_c} [\![i]\!]^{P_t} [\![i]\!]^{A_I}, & W_{(B)} &= \sum_{i=0,1} [\![1]\!]^{P_c} [\![i]\!]^{P_t} [\![i]\!]^{B_I}.
\end{align}

\subsection{Partially-dephased quantum switches}

The examples of QC-QCs presented in Fig.~2 of the extended abstract can formally be described as follows, building on the quantum switch above.
Two different bases will be considered for dephasing the various (qubit) systems: the computational $\{\ket{0/1}\}$ basis (the eigenbasis of the Pauli $Z$ operator), in which case we denote the dephasing map on system $S$ (as defined in Sec.~\ref{subsec:dephasing}) by $\Delta_S^{0/1}$; and the ``diagonal'' $\{\ket{\pm}\}$ basis (with $\ket{\pm}\coloneqq\frac{1}{\sqrt{2}}(\ket{0}\pm\ket{1})$: the eigenbasis of the Pauli $X$ operator), in which case we denote the corresponding dephasing map by $\Delta_S^\pm$.

\begin{itemize}

    \item The first example (Fig.~2, left) is one where only $P$ remains non-classical: it is obtained from the quantum switch by inputting the target state $\ket{0}^{P_t}$ (while just leaving $P_c=P$ ``open''), dephasing systems $A_I, A_O, B_I, B_O$ in the computational basis, tracing out $F_t$, and dephasing $F_c = F$ in the diagonal basis.

Its process matrix description is obtained as
\begin{align}
    \hspace{-20mm} W_1 & = \left(\Delta_{A_{IO}B_{IO}}^{0/1}\circ\Delta_{F_c}^\pm\right)\left[\Tr_{F_t}\left(\bra{0}^{P_t}W_{\mathrm{QS}}\ket{0}^{P_t}\right)\right] \notag \\
    & = \sum_{i=0,1} \left[ [\![0]\!]^{P_c} [\![0]\!]^{A_I} [\![i]\!]^{A_O} [\![i]\!]^{B_I} \id^{B_O} \frac{\id^{F_c}}{2} + [\![1]\!]^{P_c} [\![0]\!]^{B_I} [\![i]\!]^{B_O} [\![i]\!]^{A_I} \id^{A_O} \frac{\id^{F_c}}{2} \right] + X^{P_c} [\![0]\!]^{A_I} [\![0]\!]^{A_O} [\![0]\!]^{B_I} [\![0]\!]^{B_O} \frac{X^{F_c}}{2}
\end{align}
(with implicit identity operators in the first line: $\bra{0}^{P_t}W_{\mathrm{QS}}\ket{0}^{P_t} = \big(\bra{0}^{P_t}\otimes\id^{P_cA_{IO}B_{IO}F}\big)W_{\mathrm{QS}}\big(\ket{0}^{P_t}\otimes\id^{P_cA_{IO}B_{IO}F}\big)$).

    \item The second example (Fig.~2, middle) is one where only $A_I$ remains non-classical: it is obtained from the quantum switch by inputting $\ket{+,+}^{P_cP_t}$ (here the past gets fully closed), dephasing $A_O, B_I, B_O$ in the computational basis, tracing out $F_t$, and dephasing $F_c = F$ in the diagonal basis.

Its process matrix description is obtained as
\begin{align}
    \hspace{-20mm} W_2 & = \left(\Delta_{A_OB_{IO}}^{0/1}\circ\Delta_{F_c}^\pm\right)\left[\Tr_{F_t}\left(\bra{+,+}^{P_cP_t}W_{\mathrm{QS}}\ket{+,+}^{P_cP_t}\right)\right] \notag \\
    & = \frac12 \sum_{i=0,1} \left[ [\![+]\!]^{A_I} [\![i]\!]^{A_O} [\![i]\!]^{B_I} \id^{B_O} \frac{\id^{F_c}}{2} + \frac{\id^{B_I}}{2} [\![i]\!]^{B_O} [\![i]\!]^{A_I} \id^{A_O} \frac{\id^{F_c}}{2} + \left([\![i]\!]+\frac{X}{2}\right)^{A_I} [\![i]\!]^{A_O} [\![i]\!]^{B_I} [\![i]\!]^{B_O} \frac{X^{F_c}}{2} \right].
\end{align}

    \item The third example (Fig.~2, right) is one where only $A_O$ remains non-classical: it is obtained from the quantum switch by inputting $\ket{+,0}^{P_cP_t}$ (here again the past gets fully closed), dephasing $A_I, B_I, B_O$ in the computational basis, and dephasing both $F_t$ and $F_c$ ($F_tF_c = F$) in the diagonal basis.

Its process matrix description is obtained as
\begin{align}
    \hspace{-20mm} W_3 & = \left(\Delta_{A_IB_{IO}}^{0/1}\circ\Delta_{F_cF_t}^\pm\right)\left[\bra{+,0}^{P_cP_t}W_{\mathrm{QS}}\ket{+,0}^{P_cP_t}\right] \notag \\
    & = \frac12 \sum_{i=0,1} \left[ [\![0]\!]^{A_I} [\![i]\!]^{A_O} [\![i]\!]^{B_I} \frac{\id^{B_OF_t}}{2}\, \frac{\id^{F_c}}{2} + [\![0]\!]^{B_I} [\![i]\!]^{B_O} [\![i]\!]^{A_I} \frac{\id^{A_OF_t}+X^{A_O} X^{F_t}}{2}\, \frac{\id^{F_c}}{2} \right] \notag \\
    & \ \ + \frac12 \, [\![0]\!]^{B_I} [\![0]\!]^{B_O} [\![0]\!]^{A_I} \frac{\id^{A_OF_t} + Z^{A_O} \id^{F_t} + X^{A_O} X^{F_t}}{2}\, \frac{X^{F_c}}{2}.
\end{align}

\end{itemize}

\bigskip

It can be verified via semidefinite programming that $W_1, W_2, W_3$ are not QC-CCs (they don't admit a decomposition as in Eqs.~\eqref{eq:cstr_QCCC2_3}--\eqref{eq:cstr_QCCC2_0}) -- hence (since causally separable processes coincide with QC-CCs for $N=2$~\cite{Wechs_2019,Wechs2021QCQC}), despite all the dephased systems, they are causally nonseparable.

\FloatBarrier
\bibliographystyle{apsrev4-2}

\end{document}